\documentclass[12pt]{article}
\usepackage{amsfonts}
\usepackage{amsmath}
\usepackage{epsfig}

\begin{document}

\title{\textbf{How Fama Went Wrong: Measures of Multivariate Kurtosis for the
Identification of the Dynamics of a $N-$Dimensional Market}}
\author{Tanya Ara\'{u}jo$^1$\thanks{%
 \emph{Corresponding author} (tanya@iseg.utl.pt)}, Jo\~{a}o Dias$^1$,
Samuel Eleut\'{e}rio$^{2}$ and Francisco Lou\c{c}\~{a}$^1$}
\begin{small}
\date{$^{1}$ISEG, Technical University of Lisbon and UECE, R. Miguel Lupi 20,
1249-078 Lisboa, Portugal \\
$^{2}$Instituto Superior T\'{e}cnico, Av. Rovisco Pais 1049-001 Lisboa, Portugal}
\end{small}
\maketitle

\begin{abstract}
This paper investigates the common intuition suggesting
that during crises the shape of the financial market clearly differentiates
from that of random walk processes.  In this sense, it challenges the analysis
of the nature of financial markets proposed by Fama and his associates.
For this, a geometric approach is proposed in order to define the patterns of
change of the market and a measure of multivariate kurtosis is
used in order to test deviations from multinormality. The emergence of crises
can be measured in this framework, using all the available information about
the returns of the stocks under consideration and not only the index representing
the market.
\end{abstract}

\section{Introduction}

The Samuelson theory of asset prices (Samuelson, 1965) was the basis of the
Efficient Market Hypothesis (Fama, 1965, 1970, 1991), assuming that the
stock returns are approximated by a random walk and their changes are
unpredictable. Nevertheless, many scholars criticized the theoretical
assumptions of the EMH, namely its informational requisites (e.g. Grossman
and Stiglitz, 1980), detected the existence of time dependencies violating
the conditions of the theory (e.g. Lo and MacKinley, 1988), and identified
the existence of asymmetry and fat tails in the distribution of returns
(e.g. Plerou et al, 2001). Consequently, due to these and other
discrepancies, the EMH "is respected but not worshipped" in financial theory
and statistics (Pesaran, 2010: 29).

Tools for analysis of deviations from normality existed for long, e.g.
the concept of Kurtosis was introduced by Karl Pearson (1905) in order to
measure the size of the tails of a distribution as compared with those of
the normal. It was assumed, by Pearson and then by "Student" (1927), who
suggested a curious mnemonic based on the shape of animals in order to
describe non-normal distributions, that kurtosis, as well as skewness, are
common features of nature.

Fat tails are interpreted as the result of a larger part of the variance
being provoked by rare extreme events, as compared to the normal
distribution. In this sense, Pesaran discusses the effect of the recent
punctuation of crashes - such as the dot-com crash of 2000 and the general
financial crash after the subprime crisis of 2007 - and argues that periods
of bubbles and crashes deviate from market efficiency (Pesaran, 2010). This
is precisely the intuition we pursue in this paper, proposing a new approach
to the measurement of the dynamics of changes of the distributions
representing the stock returns. In the following, we describe the emergence of
crises amidst long periods of normal trading and, as we are interested in
major changes that occur at the fat tails of the distribution, we use
extensions of the concept of kurtosis to the realm of a $%
n- $dimensional object.

For this, we proceed as follows: first, we define the concepts used to
describe the dynamics of the market; the following section presents the
geometry applied to the measurement of the distortion in certain periods;
finally the next sections describe the results of the
statistics applied to relevant periods, when the market
differentiates itself from a random walk type of behavior.

\section{The identification of the dynamics of change of the market}

In the province of statistical research on financial data, the evidence of
fat tails is expressed in typical leptokurtic distributions. Recently,
Pesaran proposed an empirical verification of four indexes (S\&P, FTSE 100,
DAX, NIKKEI 225), for 2000-2009, and found evidence of kurtosis, rejecting
the null hypothesis of a normal distribution using a Jarque-Bera test
(Jarque, Bera, 1987). In the same paper, originated as a contribution to a
seminar in honor of Fama, the author mentions the historical data of the
monthly returns measured by the S\&P, as compiled by Shiller (2011): from
1871 to 2009 there is an impressive evidence of kurtosis, and even when
shorter periods are chosen, deviations from normality are typical.

As a consequence of this common empirical evidence of non-normality and
asymmetric and heavy tailed distributions of financial data, anticipated by
no less than Fama himself (Fama, 1965, also Dufour et al, 2003), several
adjusted models for skewness and kurtosis have been proposed in financial
statistics, namely in the asset pricing applications (e.g., Jarrow and Rudd,
1982, Corrado and Tsu, 1996, Brown and Robinson, 2002, Vahamaa, 2003, Iqbal
et al, 2010). In this sense, for instance Dufour, Beaulieu and Khalaf
proposed the incorporation of asymmetry of the return distribution on asset
evaluation (Dufour et al, 2003, Beaulieu et al, 2005, 2009).

The problem we address in this paper follows these lines of research but,
instead of describing the market with the recourse to a single measurement
of an index, we propose to capture the whole available empirical information
on the dynamics of a population of stocks through time, considering as a
consequence the multivariate process. There are sound reasons for this
option since, in general, the statistical experiments and empirical
approaches "are multivariate by nature" (Liu et al, 1999: 783).

By using a stochastic geometry technique, we found that the dynamics of the
S\&P500 set of stocks defines market spaces as low-dimensional entities and
that this low-dimensionality is caused by the small proportion of systematic
information present in correlations among stocks in normal periods of trade.
However, this situation changes dramatically in periods of crashes or crises.

This is verified with extensive data representing different sets of the daily
returns, namely that of the 236 S\&P500 stocks that remained in the
market from 1973 to 2009, and then verified by that of the 471 S\&P500 stocks
surviving from 2005 to 2009 (see Fig 1). These populations are used to study
the market dynamics: in both cases, in the subperiods of \emph{business-as-usual}, the geometric
object defined by the dynamics of the market approaches the spherical configuration,
typical of a Gaussian distribution; conversely, when a subperiod includes
relevant crashes, the shape of that geometric object is distorted, acquiring
prominences in some particular directions. Moreover, we found that, during
crashes, market spaces contract along their effective dimensions. In this,
we also follow a definition by R.A. Fisher (1953), establishing that,
measured in an Euclidean space, the multivariate normal errors are described
by the surface of a sphere, and suggesting that, whenever large errors
occur, the topological deviations have to be considered.

\begin{figure}[tbp]
\centering
\includegraphics[width=10cm]{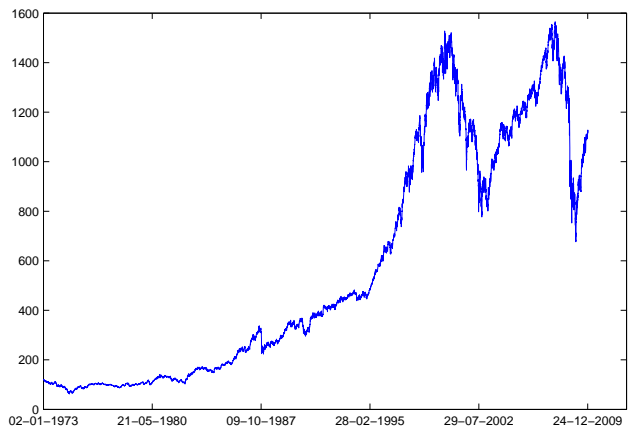}
\caption{The evolution of the S\&P500 for 1973-2009.}
\label{1.0}
\end{figure}

In order to capture the contracting and distortion effects in the market
shape, we measure multivariate kurtosis
($b_2,p$)
as presented in the next section.

\section{The measure of the distortion of the space of the market}

The strategy of measurement of the space of the financial market is simply
stated in the following terms. From the set of returns of the stocks and
their historical data of returns over the time interval, and using an
appropriate metric (Mantegna, 1999, 2000), we compute the matrix of
distances between the stocks. Considering the returns for each stock,

\begin{equation}
\begin{array}{lll}
r(k)=\log (p_{t}(k))-\log (p_{t-1}(k)) \label{1.00} &  &
\end{array}%
\end{equation}

a normalized vector

\begin{equation}
\begin{array}{lll}
\overrightarrow{\rho } (k)=\frac{\overrightarrow{r}(k)-\left\langle
\overrightarrow{r}(k)\right\rangle }{\sqrt{n\left( \left\langle
r^{2}(k)\right\rangle -\left\langle r(k)\right\rangle ^{2}\right) }} \label%
{2.00} &  &
\end{array}%
\end{equation}

is defined, where $n$ is the number of components (number of time labels) in
the vector $\overrightarrow{\rho }$. With this vector the distance between
the stocks $k$ and $l$ is defined by the Euclidean distance of the
normalized vectors.

\begin{equation}
\begin{array}{lll}
d_{kl}=\sqrt{2\left( 1-C_{ij}\right) }=\left\Vert \overrightarrow{\rho }(k)-%
\overrightarrow{\rho }(l)\right\Vert \label{3.00} &  &
\end{array}%
\end{equation}

with $C_{kl}$ being the correlation coefficient of the returns $r(k)$,$r(l)$.

\begin{equation}
C_{kl}=\frac{\left\langle \overrightarrow{r}(k)\overrightarrow{r}%
(l)\right\rangle -\left\langle \overrightarrow{r}(k)\right\rangle
\left\langle \overrightarrow{r}(l)\right\rangle }{\sqrt{\left( \left\langle
\overrightarrow{r}^{2}(k)\right\rangle -\left\langle \overrightarrow{r}%
(k)\right\rangle ^{2}\right) \left( \left\langle \overrightarrow{r}%
^{2}(l)\right\rangle -\left\langle \overrightarrow{r}(l)\right\rangle
^{2}\right) }}  \label{4}
\end{equation}

As the distance is properly defined according to the due metric axioms, it
is possible to obtain, from the matrix of distances, the coordinates for the
stocks in a Euclidean space of dimension smaller than $N$. The standard
analysis of reduction of the coordinates is applied to the center of mass
and the eigenvectors of the inertial tensor are then computed.

The same technique is also applied to surrogate (time-permuted and random) data, namely to data
obtained by independent time permutation for each stock, and these
eigenvalues are compared with those obtained from actual data in order to
identify the characteristic directions for which the eigenvalues are
significantly different. They define a reduced subspace of dimension $f$,
which carries the systematic information related to the market correlation
structure.

This corresponds to the identification of empirically constructed variables
that drive the market and, in this framework, the number of surviving
eigenvalues is the effective characteristic dimension of this economic
space ($f$). This procedure is the key for the following method, since it allows
for the consideration of populations of hundreds of stocks, given that only a
very small number of coordinates describing their distances is used in the
computation of our measures of the multivariate space.

This suggests the definition of a \textit{systematic covariance.} For
this, we denote by $\overrightarrow{z}(k)^{(f)}$ the restriction of the $k-$%
asset to the subspace $V_{f}$. and by $d_{kl}^{(f)}$ the distances
restricted to this space. Then using Eqs.(3) and (4) we may define a notion
of \textit{systematic covariance }$\sigma _{kl}^{(f)}$

\begin{equation}
\sigma _{kl}^{(f)}=\mu _{k}\sqrt{\sigma _{kk}-\overline{r}_{k}^{2}}\mu _{l}%
\sqrt{\sigma _{ll}-\overline{r}_{l}^{2}}\left( 1-\frac{1}{2}\left(
d_{kl}^{(f)}\right) ^{2}\right)  \label{2.9}
\end{equation}
where $\mu _{k}=|\overrightarrow{z}(k)^{(f)}|/|\overrightarrow{z}(k)|$ , $%
\overline{r}_{k}=\left\langle \overrightarrow{r}(k)\right\rangle $ and $%
\sigma _{kk}=\left\langle \overrightarrow{r}(k)\overrightarrow{r}%
(k)\right\rangle $ .

\bigskip

As Figure 2 clearly indicates, in
periods of normal trade the spherical configuration is maintained, similar
to that of surrogate data, whereas in periods of turbulence new shapes
emerge.

\begin{figure}[tbp]
\centering
\includegraphics[width=10cm]{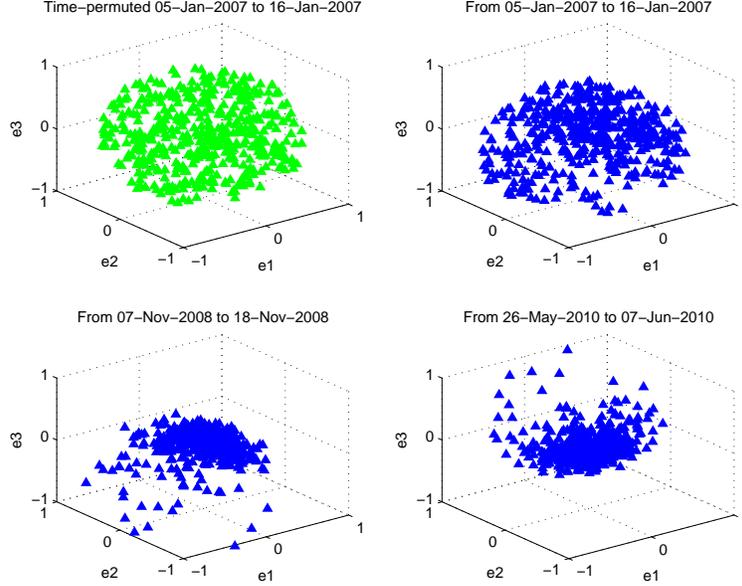}
\caption{The market spaces along their main 3 dimensions: each point describes
a firm, for surrogate data and \emph{business-as-usual} periods (the upper plots)
correspond to a spherical shape while in periods of crises (the plots in the
bottom) proeminences in the market shape are obvious.}
\label{2.0}
\end{figure}

The plots of Figure 2 show that while in the first days of January 2007
there is no relevant difference in relation to surrogate (time-permuted) data, new
shapes emerge in March 2007 as well as in November 2008, two periods of
strong turbulence in the financial markets, indicated by a distortion in the
shape of the sphere.

\bigskip

\section{A Test Based on Multivariate Kurtosis}

\medskip

Given this evidence of contraction and distortion of the market space in different
periods, we proceed to test deviation from normality. The computation of
univariate measures and tests of deviations of general indexes from normality
is a common procedure (e.g., Pesaran, 2010). Instead, we propose to consider
the available information on the richest detail of a large population of stocks.
For this, we recur to heuristic concepts and a test of multivariate kurtosis
($b_2,p$), such as proposed by Mardia (1970). In this case, multivariate kurtosis
is defined as

\begin{equation}
\begin{array}{lll}
b_2,p(t)=\frac{1}{N} \Sigma _{i}[(z_{i}(t)-\overline{z})(\sigma^{(f)})^{-1}(z_{i}(t)-\overline{z})]^2 \label{6.00} &  &
\end{array}%
\end{equation}

\medskip where $\sigma ^{(f)}$ is the systematic covariance, $p$ is the number of
variables and $N$ the number of observations. In the calculation of $\sigma ^{(f)}$ 
$f=6$ since six was found (Ara\'{u}jo and Lou\c{c}\~{a}, 2007,
2008) to be the number of effective dimensions of the S\&P500 market space.

Although the statistical properties of tests of multivariate normality are
not as established as those applied to univariate normality (Mardia et al,
1979, Gnanadesikan, 1997), a large body of literature was built in the last
decades on the topic. Mardia (1970, 1974, 1983), using Arnold's results
(Arnold, 1964) proposed the affine invariant measures as previously discriminated,
established their asymptotic distributions and formulated tests for the null of
multivariate normality. Different scholars discussed the limit distributions of Mardia's
tests (Schwager, 1985, Baringhaus and Henze, 1992, Kariya and George, 1995,
Zhao and Konichi, 1997) and investigated their consistency (Baringhaus and Henze,
1988, Henze, 1994). Others, as Koziol (1982, 1983, 1993, 2005),
Srivastava (1984) and Henze (1990), proposed alternative
approaches to test multivariate normality.

Several authors surveyed the available procedures and tests of multivariate
normality and identified more than fifty alternatives, although only some
qualified as generally accepted methods (Horswell and Looney, 1992, Looney,
1995, Mecklin and Mundfrom, 2004, 2005, Szekely and Rizzo, 2005, Farrell et
al, 2006). Through the comparison of the power of the different tests, using
extensive Monte Carlo simulations, some of these authors argued that the
Mardia tests have low power (Farrell et al, 2006) in particular against the
BHEP test proposed by Henze and his collaborators (Mecklin and Mundfrom,
2004, 2005), whereas others obtained an opposite conclusion, favoring the
Mardia test (Dufour et al, 2003). Bai and Ng, considering the fact that the
sampling distributions of these coefficients is not well known for serially
correlated data, proposed a strategy of generalization of the Jarque-Bera
test in order to account for these problems (Bai, Ng, 2005, also Doornik
and Hansen, 2008).

Considering the stationarity of our series of returns, by construction, we
are nevertheless confronted with evidence of serial and cross correlation.
It is well known that daily returns tend to be negatively serially
correlated, that their statistical significance tends to be greater in
periods of unrest and that their cross correlation increases with volatility
(Pesaran, 2010). A novel approach to deal with this problem, considering
serial correlation in the residuals of overlapping observations, was
proposed recently by Pesaran and his colleagues, using a new version of a
seemingly unrelated regression equations based estimation (Pesaran et al,
2011). Furthermore, there is evidence of cross correlation, which has been
rarely discussed in the framework of multivariate analysis, with some
exceptions (e.g., Richardson and Smith, 1993).

Considering these suggestions, we checked our data applying the Mardia
measure to the stocks in our population, for each period, in order to
diagnose deviations from normality and to describe the dynamics of the market
in different periods. In each case we proceeded to systematic comparisons
with the measures of series
of random data obtained from a Gaussian distribution with the same average
and variance as in our population.

Mardia's test of multivariate normality is performed in order to determine
if the null hypothesis of multivariate normality is a reasonable
assumption regarding the population distribution of a random sample.

Mardia proved that, under the null hypothesis, the statistics

\begin{equation}
\begin{array}{lll}
t2=\frac{b_2,p(t)-(p^2+2p)}{\sqrt{\frac{8p^2+16p}{N}}} &  &
\end{array}
\end{equation}

is asymptotically distributed as a standard normal.
Since our aim is not only to test
the data but essentially to distinguish the periods of \emph{business-as-usual}
from the periods of crises and, in that sense, to test the distortions in
the dynamics of the markets, we replace the expected value and standard deviation used in Mardia's
standardization with the empiric counterparts as obtained from the observed values of the statistics in a \emph{business-as-usual}
period. In this sense, we consider \emph{business-as-usual} periods 1973-1995 for the
longer series and January 2005 to June 2007 for the shorter series. The modified
statistics is then

\begin{equation}
\begin{array}{lll}
g(t)=\frac{b_2,p(t)-\overline{b_2,p(t)}}{\widehat{\sigma}(b_2,p(t))} &  &
\end{array}
\end{equation}

where $\overline{b_2,p(t)}$ and $\widehat{\sigma}(b_2,p(t))$ are the estimated
values of, respectively, the mean and the standard deviation. In this case, our
variables are the six coordinates identifying the relevant dimensions, which
represent the relevant information about the market. This is a robust result,
unaltered even when other dimensions are considered, confirming our previous
result indicating that such dimensions essentially represent noise. Even if
the Mardia test - or other tests on multivariate normality - cannot be
applied to a very large number of variables, since the asymptotic properties are
not known for those cases, this strategy allows both for considering a large
population (236 and 471 firms) and to test the described dynamics of the
market, considering all the available information and not just a single index
averaging through the market.

\section{Results and Discussion}

\medskip

In previous empirical work, we found a robust result: markets of different
sizes, ranging from 70 to 424 stocks across different time windows (from one
year to 35 years), and using different market indexes for different markets,
may be described by six effective dimensions (Ara\'{u}jo and Lou\c{c}\~{a}, 2007,
2008). A striking characteristic of the data is also that, whenever the market
suffers a crash, there is a contracting effect provoking a clear distortion in
the dominant directions of the market space. If the volume expands whenever the
cloud of points represents a situation of \emph{business-as-usual} and the market space
is similar to that of a random universe, whenever a crisis occurs the volume of the
geometric object severely contracts, leading to the emergence of
characteristic and distorted shapes.

\begin{figure}[tbp]
\centering
\includegraphics[width=10cm]{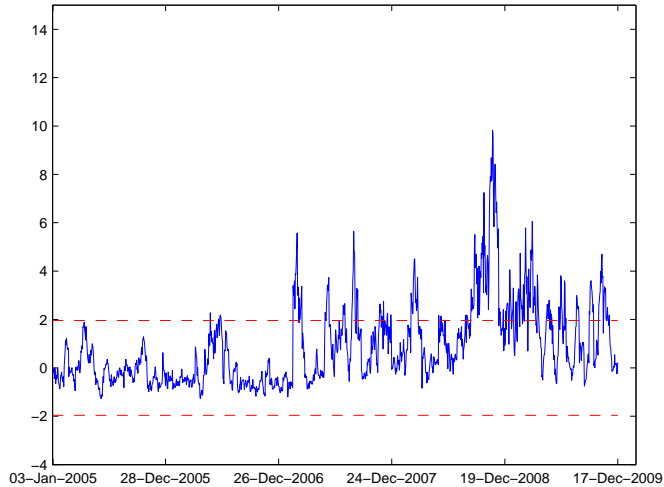}
\caption{The evolution of g(t) from 2005 to 2009.}
\label{3.0}
\end{figure}

Figure 3 describes the evolution of g(t)
for 2005-2009 and Figure 4 the same statistic
for the larger period of 1973-2009, and the typical limits of the interval
corresponding to a level of significance of 5\% are indicated. The hypothesis of
normality is clearly rejected in the periods of turbulence, as expected.

\begin{figure}[tbp]
\centering
\includegraphics[width=10cm]{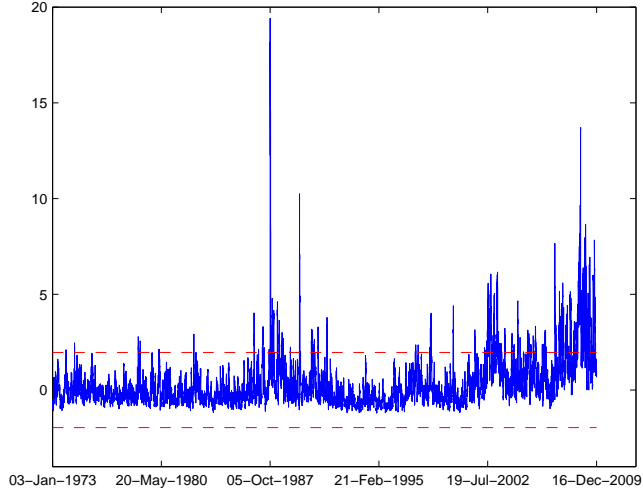}
\caption{The evolution of g(t) from 1973 to 2009.}
\label{4.0}
\end{figure}

In Figure 5 the results of the test for random data with the same standard
deviation and expected value as the true data are presented, highlighting by comparison
the presence of structure in the market described by our data.

\begin{figure}[b]
\centering
\includegraphics[width=6cm]{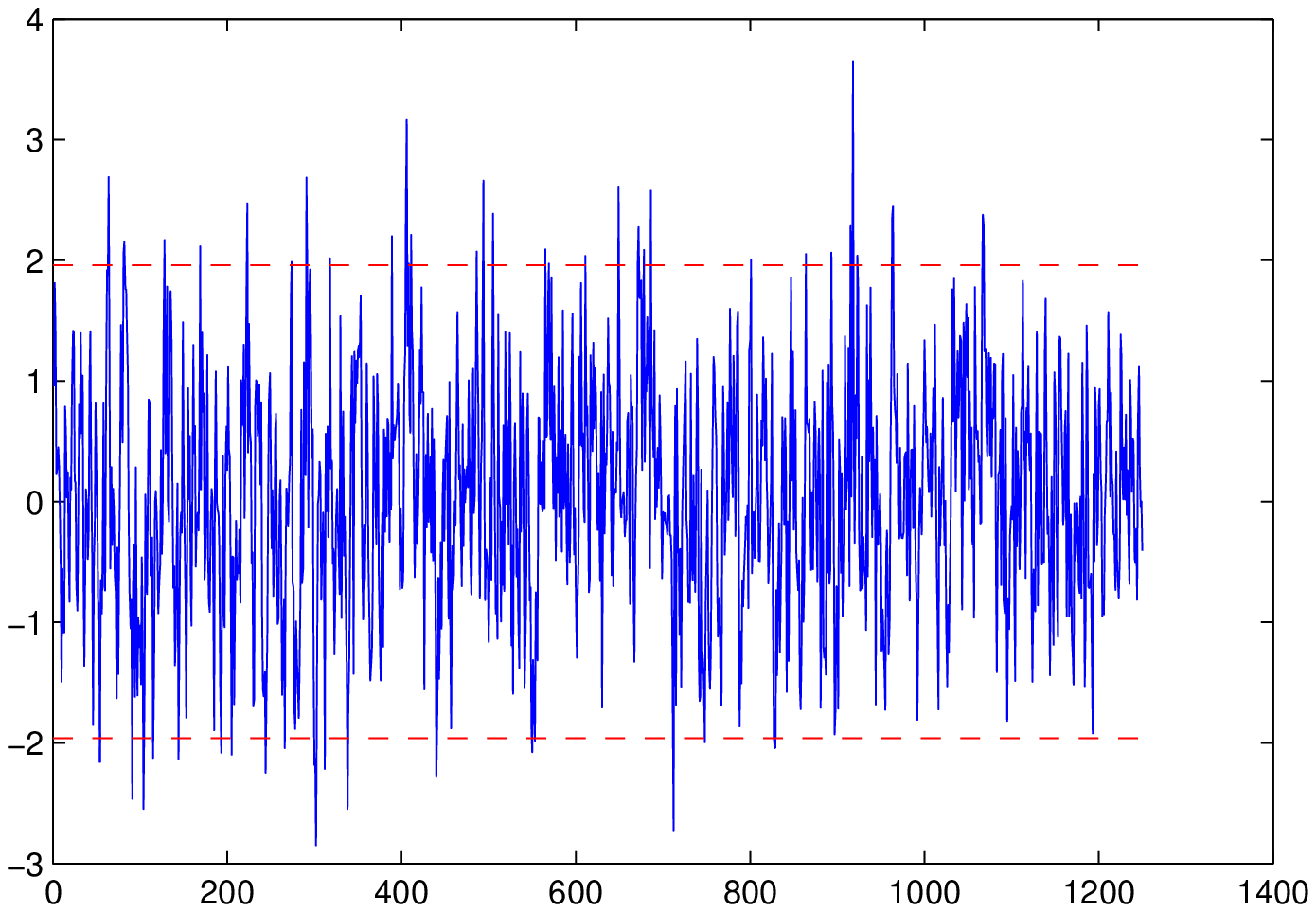}\includegraphics[width=6cm]{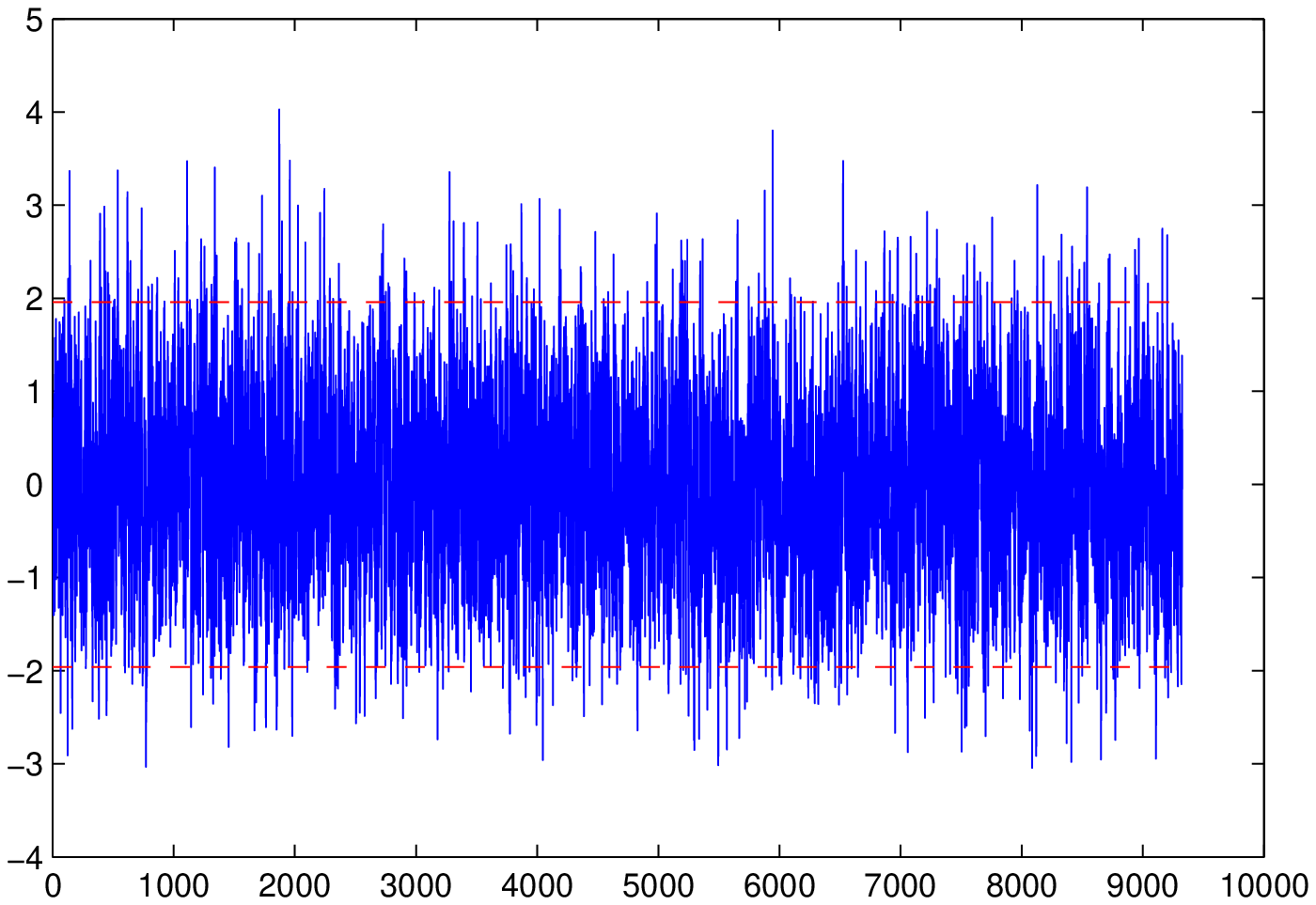}
\caption{The evolution of g(t)
corresponding to random data with the same mean and stardard deviation as in Figure 3 (l.h.s.) and Figure 4 (r.h.s.)}
\label{5.0}
\end{figure}

\section{Conclusion}

The multivariate analysis has rarely been applied to stock market series. In this
paper, we propose a novel approach to that application, which permits the consideration
of the evolution of a large population of stocks for different periods and describes
the deformations in the geometry of the market, while discussing some of the technical
difficulties involved in this enterprise. This is possible since the evolution
of the distances among all the firms is encapsulated by a small manifold of coordinates
in the market space. Although this presents some statistical difficulties we are addressing
in the current research, this avenue favors a new approach to the multidimensional objects
constructed by the dynamics of complex markets and interactions among many firms
and decisions.

As it was previously found, during periods of financial turbulence the shape of the
market changes dramatically, whereas in periods of normal business it resembles the
spherical form typical of a random distribution. Our data confirms those results.
A moment, kurtosis, is used in order to measure the distortions of the distribution,
and the results highlight how the distances among stocks contract in periods of unrest.
Given this, the hypothesis of well behaved random distribution of the stock returns can
be challenged and should be challenged. It does not correspond to the state of nature
during a crisis or an episode of turbulence in the financial markets we investigated.

\section*{Acknowledgement}

This work has benefited from partial
financial support from the Funda\c{c}\~{a}o  para a Ci\^{e}ncia e a Tecnologia-FCT, under
the 13 Multi-annual Funding Project of UECE, ISEG, Technical University of
Lisbon.

\section{References}

\begin{itemize}
\item Ara\'{u}jo T., Lou\c{c}\~{a}, F. 2007.The Geometry of Crashes - A
Measure of the Dynamics of Stock Market Crises. Quantitative Finance\textbf{%
\ 7(1)}: 63-74.

\item Ara\'{u}jo T., Lou\c{c}\~{a}, F. 2008. The Seismography of Crashes in
Financial Markets. Physics Letters A \textbf{372-4}: 429-34.

\item Arcones, M. 2007. Two Tests for Multivariate Normality Based on the
Characteristic Function. Mathematical Methods of Statistics \textbf{16(3)}:
177-201.

\item Arnold, H.J. 1964. Permutation Support for Multivariate Techniques.
Biometrika \textbf{51}: 65-70.

\item Averous, J., Meste, M. 1997. Skewness for Multivariate Distributions:
Two Approaches. Annals of Statistics \textbf{25(5)}: 1984-97.

\item Bai, J., Ng, S. 2005. Tests for Skewness, Kurtosis and Normality for
Time Series Data. Journal of Business and Economic Statistics \textbf{23(1)}%
: 49-60.

\item Baringhaus, L., Henze, N. 1988. A Consistent Test for Multivariate
Normality. Metrika \textbf{35}: 339-48.

\item Baringhaus, L., Henze, N. 1992. Limit Distributions for Mardia's Measure of Multivariate
Skewness. Annals of Statistics \textbf{20(4)}: 1889-1902.

\item Beaulieu, M.-C., Dufour, J.-M., Khalaf, L. 2005. Exact Multivariate
Tests of Asset Pricing Models with Stable Asymmetric Distributions. In
Numerical Methods in Finance, edited by Breton, M., Ben Ameur, H., pp.
173-91. Dordretch: Kluwer.

\item Beaulieu, M.-C., Dufour, J.-M., Khalaf, L. 2009. Finite Sample Multivariate Tests of Asset Pricing Models with Coskewness. Computational Statistics and Data Analysis \textbf{53}: 2008-21.

\item Brown, C., Robinson, D. 2002. Skewness and Kurtosis Implied by Option
Prices: A Correction, Journal of Financial Research \textbf{25}: 279-82.

\item Corrado, C., Tsu, T. 1996. Skewness and Kurtosis in S\&P500 Index
Returns Implied by Option Prices, Journal of Financial Research \textbf{19}:
175-92.

\item Doornik, J., Hansen, H. 2008. An Omnibus Test for Univariate and
Multivariate Normality. Oxford Bulletin of Economics and Statistics \textbf{%
70}: 927-39.

\item Dufour, J.-M., Khalaf, L., Beaulieu, M.-C. 2003. Exact
Skewness-Kurtosis Tests for Multivariate Normality and Goodness-of-fit in
Multivariate Regressions with Application to Asset Pricing Models. Oxford
Bulletin of Economics and Statistics \textbf{65}: 891-906.

\item Eonomoto, R., Okamoto, N., Seo, T. ND. On the Distribution of Test
Statistic Using Srivastava's Skewness and Kurtosis.

\item Fama, E.F. 1965. The Behaviour of Stock Prices. Journal of Business
\textbf{60}: 401-24.

\item Fama, E.F. 1970. Efficient Capital Markets: A Review of Theory and Empirical
Work, Journal of Finance \textbf{25(2)}: 383-420.

\item Fama, E.F. 1991. Efficient Capital Markets: II, Journal of Finance \textbf{46(5)}%
: 1575-617.

\item Fama, E.F., French, K.R. The CAPM: Theory and Evidence. Working Paper,
Center for Research in Security Prices, University of Chicago.

\item Farrell, P., Salibian-Barrera, M., Naczk, K. 2006. On Tests for
Multivariate Normality and Associated Simulation Studies. Journal of
Statistical Computation and Simulation \textbf{0}:1-14.

\item Fisher, R.A. 1953. Dispersion on a Sphere. Proceedings of the Royal
Society \textbf{217}: 295-305.

\item Gnanadesikan, R. 1997. Methods for Statistical Data Analysis of
Multivariate Observations. New York: Wiley.

\item Gosset, W. 1927. Errors on Routine Analysis. Biometrika \textbf{19(1-2)%
}: 151-64.

\item Grossman, S.J., Stiglitz, J. 1980. On the Impossibility of
Informationally Efficient Markets. American Economic Review \textbf{70},
393-408.

\item Henze, N. 1994. On Mardia's Kurtosis Test for Multivariate Normality.
Communications in Statistics - Theory and Methods \textbf{23}: 1031-45.

\item Henze, N. 2002. Invariant Tests for Multivariate Normality: A Critical Review.
Statistical Papers \textbf{43}: 467-506.

\item Henze, N., Penrose, M.D. 1999. On the Multivariate Runs Test. Annals
of Statistics \textbf{27(1)}: 290-8.

\item Henze, N., Wagner, T. 1997. A New Approach to the BHEP tests for
Multivariate Normality. Journal of Multivariate Analysis \textbf{62}: 1-23.

\item Henze., N., Zirkler, B. 1990. A Class of Invariant and Consistent
Tests for Multivariate Normality. Communications in Statistics - Theory and
Methods \textbf{19}: 3595-617.

\item Horswell, R., Looney, S. 1992. A Comparison of Tests for Multivariate
Normality that are Based on Measures of Multivariate Skewness and Kurtosis.
Journal of Statistical Computation and Simulation \textbf{42}: 21-38.

\item Iqbal, J., Brooks, R., Galagedera, D. 2010. Multivariate Tests of
Asset PRicing: Simulation Evidence from an Emerging Market \textbf{20}:
381-95-

\item Jarque, C.M., Bera, A.K. 1980. Efficient Tests for Normality,
Homoskedasticity and Serial Independence of Regression Residuals. Economic
Letters \textbf{12}: 255-9.

\item Jarque, C.M., Bera, A.K. 1987. A Test for Normality of Observations and Regression Residuals.
International Statistical Review \textbf{55}: 163-72.

\item Jarrow, R., Rudd, A. 1982. Approximate Option Valuation for Arbitrary
Stochastic Processes. Journal of Financial Economics \textbf{10}, 347-69.

\item Kariya, T., George, E. 1995. LBI Tests for Multivariate Normality in
Curved Families and Mardia's Test. Sankhya: The Indian Journal of Statistics
\textbf{57(3), series A}: 440-51.

\item Klar, B. 2002. A Treatment of Multivariate Skewness, Kurtosis and
Related Statistics. Journal of Multivariate Analysis \textbf{83(1)}: 141-65.

\item Koziol, J. 1978. Exact Slopes of Certain Multivariate Tests of
Hypotheses. Annals of Statistics \textbf{6(3)}: 546-58.

\item Koziol, J. 1982. A Class of Invariant Procedures for Assessing Multivariate
Normality. Biometrika \textbf{69(2)}: 423-7.

\item Koziol, J. 1983. On Assessing Multivariate Normality. Journal of the Royal
Statistical Society. Series B, Methodological \textbf{45(3)}: 358-61.

\item Koziol, J. 1993. Probability Plots for Assessing Multivariate Normality. Journal
of the Royal Statistical Society. Series D, Statistician \textbf{42(2)}:
161-73.

\item Liu, R., Parelius, J., Singh, K. 1999. Multivariate Analysis by Data
Depth: Descriptive Statistics, Graphics and Inference. Annals of Statistics
\textbf{27(3)}: 783-840.

\item Lo, A., MacKinley. A.C. 1988. Stock Prices do not Follow Random Walks:
Evidence from a Simple Specification Test. Review of Financial Studies
\textbf{1}, 41-66.

\item Looney, S.W. 1995. How to Use Tests for Univariate Normality to Assess
Multivariate Normality. American Statistician \textbf{49(1)}: 64-70.

\item Mantegna R.N. 1999. Hierarchical structure in financial markets.
European Physics Journal\textbf{\ B11}: 193-7.

\item Mantegna, R.N., Stanley, H.E. 2000. An Introduction to Econophysics:
Correlations and Complexity in Finance. Cambridge, Cambridge University
Press.

\item Mardia, K.V. 1970. Measures of multivariate skewness and kurtosis with
applications. Biometrika \textbf{57}: 519-30.

\item Mardia, K.V. 1974. Applications of Some Measures of Multivariate Skewness and
Kurtosis in Testing Normality and Robustness Studies. Sankhya \textbf{B36}:
115-28.

\item Mardia, K.V. 1980. Tests of Univariate and Multivariate Normality. Handbook of
Statistics, edited by Krishnaiah, volume 1, 297-221. Amsterdam: North
Holland.

\item Mardia, K.V., Kanazawa, M. 1983. The Null Distribution of Multivariate
Kurtosis. Communications in Statistical Simulation and Computation \textbf{12%
}: 569-76.

\item Mardia, K.V., Kent, J., Bibby, J. 1979. Multivariate Analysis. New
York: Academic Press.

\item Mardia, K.V., Zemroch, P.J. 1975. Algorithm AS 84: Measures of
Multivariate Skewness and Kurtosis. Journal of the Royal Statistical Society
\textbf{24(2), series C}: 262-5.

\item Mecklin, C., Mundfrom, D. 2004. An Appraisal of Bibliography of Tests
for Multivariate Normality. International Statistical Review \textbf{72}:
123-38.

\item Mecklin, C., Mundfrom, D. 2005. A Monte Carlo Comparison of the Type I and Type
II Error Rates of Tests of Multivariate Normality. Journal of Statistical Computation and
Simulation \textbf{75}: 93-107.

\item Pearson, K. 1905. Das Fehlergesetz und Seine Verallgemeiner-ungen
Durch Fechner und Pearson, A Rejoinder. Biometrika, \textbf{4}: 169-212.

\item Pesaran, M.H. 2010. Predictability of Asset Returns and the Efficient
Market Hypothesis. IZA Discussion Papers, Institute for the Study of Labor.

\item Pesaran, M.H., Pick, A., Timmermann, A. 2011. Variable Selection,
Estimation and Inference for Multi-Period Forecasting Problems. Journal of
Econometrics \textbf{64(1)}: 173-87.

\item Plerou, V., Gopikrishnan, P., Gabaix, X., Amaral, L.A., Stanley, H.E.
2001. Price Fluctuations, Market Activity and Trading Volume. Quantitative
Finance \textbf{1}: 262-9.

\item Richardson, M., Smith, T. 1993. A Test for Multivariate Normality in
Stock Returns. Journal of Business \textbf{66}: 295-321.

\item Samuelson, P. 1965. Proof that Properly Anticipated Prices Fluctuate
Randomly. Industrial Management Review Spring \textbf{6}: 41-9.

\item Schwager, S.J. 1985. Multivariate Skewness and Kurtosis. Encyclopaedia
of Statistical Sciences \textbf{6}: 122-5. New York; Wiley.

\item Shiller, R. 2011. http://www.econ.yale.edu/~shiller/data.htm.

\item Srivastava, M.S. 1984 A Measure of Skewness and Kurtosis and a
Graphical Method for Assessing Multivariate Normality. Statistical and
Probability Letters \textbf{2}: 263-7

\item Szekely, G., Rizzo, M. 2005. A New Test for Multivariate Normality.
Journal of Multivariate Analysis \textbf{93}: 58-80.

\item Vahamaa, S. 2003. Skewness and Kurtosis Dajusted Black-Scholes Model:
A Note on Hedging Performance, Finance Letters \textbf{1(5)}: 6-12.

\item Vilela-Mendes, R., Ara\'{u}jo, T. and Lou\c{c}\~{a}, F. 2003.
Reconstructing an Economic Space from a Market Metric. Physica A \textbf{323}%
: 635-50.

\item Zhao, Y., Konishi, S. 1997. Limit Distributions of Multivariate
Kurtosis and Moments under Watson Rotational Symmetric Distributions.
Statistical and Probability Letters \textbf{32(3)}: 291-9.
\end{itemize}

\end{document}